\documentclass[sigconf]{acmart}

\usepackage{booktabs} 
\usepackage{balance}
\usepackage{graphics}
\usepackage{amsfonts}
\usepackage{array}
\usepackage[multiple]{footmisc}
\usepackage{amsmath}
\usepackage{etoolbox}
\usepackage{enumitem}
\usepackage{comment}

\newbool{showComments}
\boolfalse{showComments}    
\ifbool{showComments}{%
\newcommand{\bc}[1]{\textcolor{blue}{\textbf{Bhawana:} #1}}

}{
\newcommand{\bc}[1]{}
\newcommand{\ua}[1]{}
\newcommand{\sj}[1]{} 
\newcommand{\cm}[1]{}

}

\AtBeginDocument{%
  }

\begin{document}

\title{NeckCare: Preventing Tech Neck using Hearable-based Multimodal Sensing}

\author{Bhawana Chhaglani}
\authornote{This work was done when the author was on an internship at Dolby Laboratories.}
\affiliation{%
  \institution{Univeristy of Massachusetts Amherst}
  \city{Amherst}
 \country{USA}
}
\email{bchhaglani@umass.edu}

\author{Alan Seefeldt}
\affiliation{
\institution{Dolby Laboratories}
\city{San Francisco}
\country{USA}
}
\email{ajs@dolby.com}




\renewcommand{\shortauthors}{Chhaglani et al.}

\begin{abstract}

Tech neck is a modern epidemic caused by prolonged device usage and it can lead to significant neck strain and discomfort. This paper addresses the challenge of detecting and preventing tech neck syndrome using non-invasive ubiquitous sensing techniques. We present NeckCare, a novel system leveraging hearable sensors, including IMUs and microphones, to monitor tech neck postures and estimate distance form screen in real-time. By analyzing pitch, displacement, and acoustic ranging data from 15 participants, we achieve posture classification accuracy of 96\% using IMU data alone and 99\% when combined with audio data. Our distance estimation technique is millimeter-level accurate even in noisy conditions. NeckCare provides immediate feedback to users, promoting healthier posture and reducing neck strain. Future work will explore personalizing alerts, predicting muscle strain, integrating neck exercise detection and enhancing digital eye strain prediction.
\end{abstract}

\ccsdesc[500]{Human-centered computing~Mobile devices}
\ccsdesc[300]{Human-centered computing~Collaborative and social computing devices}

\keywords{Health Sensing, Tech Neck, Multimodal Sensing, Hearables}

\pagestyle{plain}

\pagenumbering{gobble}



\maketitle

\section{Introduction}

\begin{figure}[t]
    \centering
    \includegraphics[width=0.7\columnwidth]{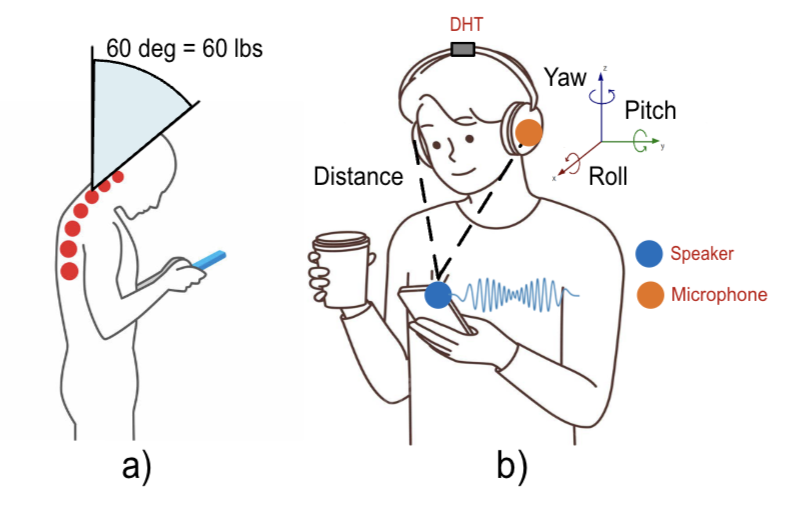}
    \vspace{-0.2in}
    \caption{a)The weight of strain that causes Tech Neck b)\textit{NeckCare} Idea  }
    \label{fig:vision}
\end{figure}

The pervasive use of digital devices has led to a widespread condition known as tech neck or text neck syndrome, characterized by neck pain and strain resulting from prolonged forward head posture or hunching. When we are working on a computer or looking down at your phone, the muscles in the back of the neck have to contract to hold your head up. The more we look down, the more the muscles have to work to keep your head up. These muscles can get overly tired and sore from looking down at our smartphones, computers, or tablets all day. Studies indicate that maintaining a 60-degree head tilt can exert pressure equivalent to 60 lbs on the neck and spine, significantly impacting musculoskeletal health \cite{website:techneck}. Prolonged use of devices not only leads to tech neck, but also causes eye strain. Shorter digital screen distance and a constant convergence can also lead to digital eye strain (DES) \cite{kaur2022digital}.
The long term effects of Tech neck include pain in shoulders and upper back, headaches/ dizziness, postural changes, pain or dysfunction in the jaw, tingling or numbness or weakness in the hands, breathing problems, and loss of lung capacity. 
With approximately 75\% of the global population spending hours daily with their heads flexed forward \cite{bottaro2022association}, there is an urgent need for effective solutions to mitigate this issue. According to a study, 73\% of university students and 64.7\% of people who work from home have neck or back pain and 39.2\% of them admit to being less productive due to neck or low back pain \cite{tsantili2022text}. Children's musculoskeletal systems are still developing, making them more susceptible to postural issues and long-term damage from poor habits. Thus, there is a need for prevention and management of tech neck by tracking and correcting the posture when using smartphones, tablets, and laptops. Since these issues (tech neck and DES) have arised due to technology, we aim to use technology to address them.\\
Various approaches have been proposed to monitor and correct user posture, including vision-based, pressure sensor-based, audio-based, and inertial measurement units (IMU)-based systems. However, these methods often face challenges such as privacy concerns, environmental dependency, and the need for external devices. Vision-based systems, for instance, are computationally expensive and sensitive to line-of-sight issues \cite{chen2019sitting}, while pressure sensor-based systems require retrofitting existing furniture \cite{tsai2023automated}. Commercial IMU-based wearables require placing them at the back of neck \cite{website:upright}. 
We aim to design a privacy-preserving system to continuously monitor user posture (distance from screen and neck angle) and provide real-time feedback. We propose a posture monitoring solution that is non-invasive, independent of environment, and does not require carrying an additional device. Our system is designed around hearables as they are ubiquitous, socially accepted, and most people wear them while working on smartphones or laptops.\\
In this paper, we present NeckCare, a novel system that leverages the sensors embedded in hearables to monitor neck posture and distance fr om the screen. 
By combining IMU and audio sensing on earbuds, our system provides a comprehensive solution for preventing tech neck. 
We use features from IMU and microphones to detect distance from the screen and important postures that cause tech neck. We provide feedback to the users to correct their posture or take breaks if they are in incorrect postures for long periods of time. 
To this end, we make the following key contributions:
\begin{itemize}
\item We develop a novel hearable-based system for real-time neck posture monitoring.
    \item By combining IMU data for neck angle detection with audio-based distance estimation, NeckCare provides a more holistic view of user posture. 
    \item We propose a method for estimating screen distance using acoustic ranging with mm-level accuracy and an algorithm for classifying 5 critical neck postures with 99\% accuracy. 
     \item We collect a robust dataset from multiple participants to train and evaluate our system, including noisy conditions and head movements. 
\end{itemize}




\section{Background and Related Work}~\label{sec:related}
{\small
\begin{table*}[h!]
\centering
\begin{tabular}{| m{3cm} | m{6cm} | m{6cm} |}
\hline
\textbf{Approach} & \textbf{Advantages} & \textbf{Disadvantages} \\
\hline
\textbf{Vision-Based \cite{tsai2023automated,kapoor2022light}} & Effective in identifying fine-grained sitting postures & Privacy concerns, computationally expensive, sensitive to line-of-sight (LoS) issues \\
\hline
\textbf{Pressure Sensor-Based \cite{chen2019sitting,fragkiadakis2019design}} & Low cost sensors; non-intrusive & Requires retrofitting furniture, location-dependent \\
\hline
\textbf{RF/Wireless Techniques \cite{li2021sitsen}} & Non-intrusive & Requires RFID Reader, environment dependency, Sensitive to environmental changes \\
\hline
\textbf{Audio-Based \cite{qu2023sitpaa,bi2024smartsit}} & Leverages already present speakers and microphones & Sensitive to environmental changes, limited to sitting postures, provides limited information \\
\hline
\textbf{IMU-Based \cite{luo2023skin,website:upright}} & Accurate, environment-independent & Requires wearing an external device, can be intrusive and inconvenient \\
\hline
\textbf{NeckCare (IMU+Audio)} & Comprehensive, non-invasive, privacy-preserving, environment-independent, no additional devices required & Requires users to wear hearables at all times\\
\hline
\end{tabular}
\caption{Comparison of different posture monitoring approaches}
\label{table:comparison}
\end{table*}
}
This section covers the existing research landscape in detecting neck or sitting postures. 
Vision-based posture recognition systems \cite{chen2019sitting} utilize computer vision techniques to identify sitting postures. These techniques take input frames from any regular camera, extract body keypoints (landmarks) from the images, and detect whether the posture is good or bad \cite{kapoor2022light}. While effective, these systems often raise privacy concerns and are computationally expensive. Additionally, they are sensitive to line-of-sight (LoS) issues, which can limit their applicability in real-world scenarios.
Pressure sensor-based techniques
\cite{tsai2023automated}, rely on pressure sensors embedded in seating arrangements to detect posture. This includes acquiring the pressure distribution of a sitting person with a number of piezoresistive sensors placed on a seat \cite{fragkiadakis2019design}. 
These systems require retrofitting existing furniture and are highly dependent on the environment, which can be a significant limitation for widespread adoption.
Audio-based posture recognition \cite{qu2023sitpaa,chhaglani2023cocoon,bi2024smartsit}, uses acoustic signals to determine sitting postures. While innovative, these systems are sensitive to environmental changes and typically only work for sitting postures. They also provide limited information compared to other sensing modalities.
IMU-based systems \cite{luo2023skin}, use inertial measurement units to monitor neck posture. These systems often require the user to wear an external device, which can be intrusive and inconvenient. However, they offer a non-invasive and environment-independent solution for posture monitoring.
Our work leverages a combination of IMU and audio sensing to monitor neck postures and distance from the screen. By integrating these modalities, we aim to overcome the limitations of individual approaches, providing a comprehensive and non-invasive solution for preventing tech neck. Exisitng work FaceOri \cite{wang2022faceori} tracks head distance and head orientation in relation to a device to feed smarter device interactions using similar idea of combining microphones and IMU sensors. However, this work focuses on applications like activity and gesture sensing and context-aware and attentive user interfaces. We leverage similar techniques for prevention of tech neck and eye strain. Apple health uses TrueDepth camera-based sensing to generate alerts for eye strain when users are too close to the screen \cite{AppleScreenDistance}. Our system is designed to be privacy-preserving, independent of the environment, and does not require users to carry additional devices. They system can accurately detect postures as well as DES when users are using devices, addressing multiple tech-related health issues.


\section{Approach and Challenges}
\label{sec:collaboration}
\begin{figure*}[t]
    \centering
    \includegraphics[width=0.98\textwidth]{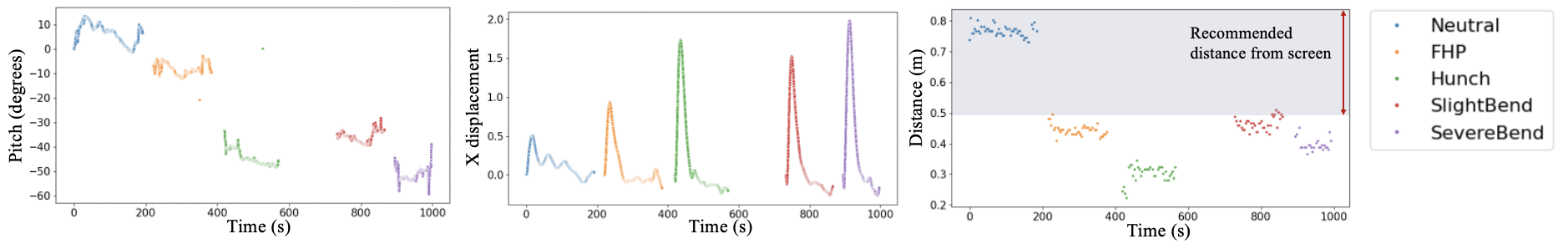}
    \caption{Sensor readings for different postures: pitch is important indicator of neck flexion angle, displacement signifies movement of head, and acoustic ranging measures distance to the screen.}
    \label{fig:key}
\end{figure*}
\begin{figure}[t]
    \centering
    \includegraphics[width=0.4\columnwidth]{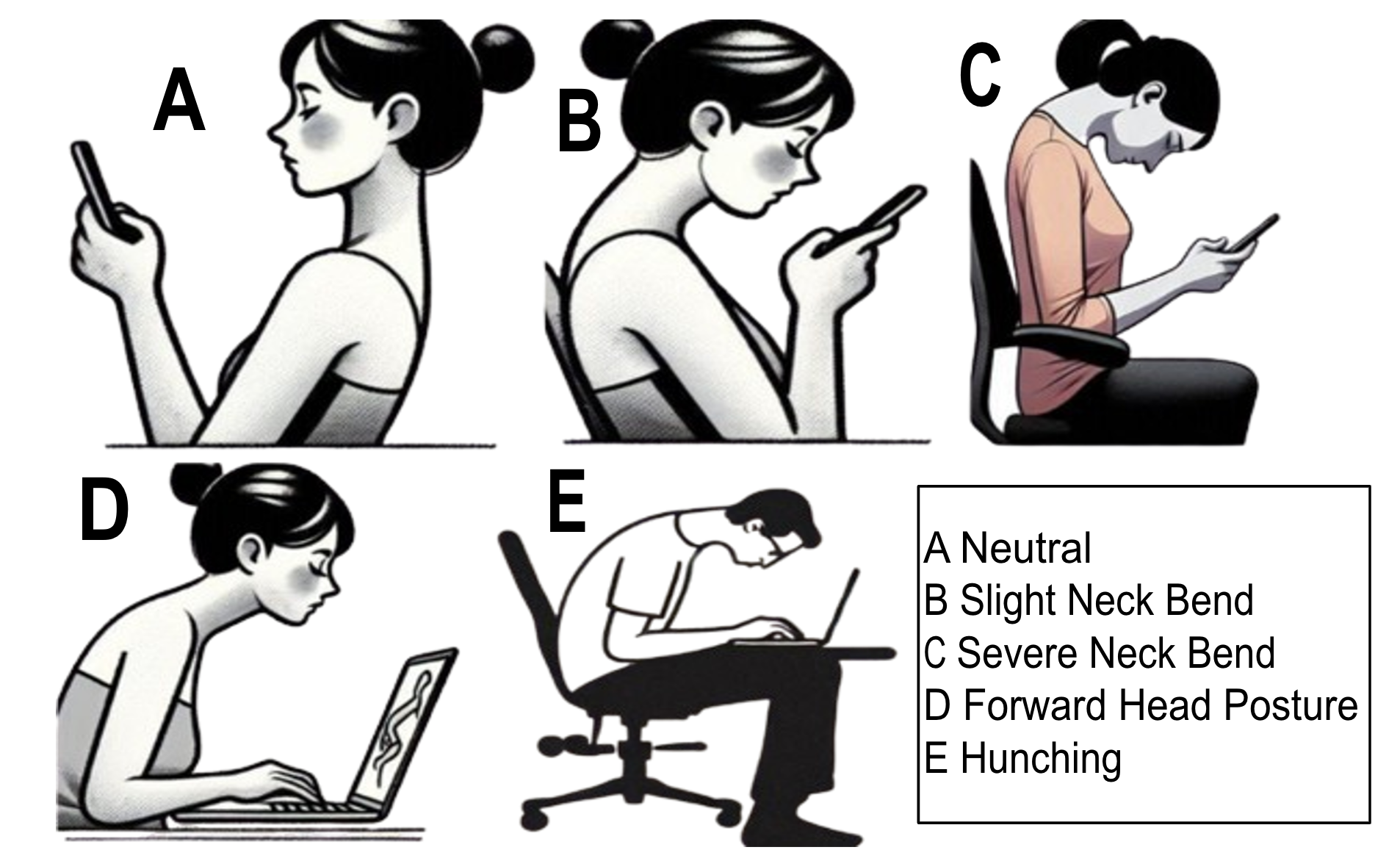}
    \caption{Important Postures for Tech Neck}
    \label{fig:postures}
\end{figure}

This section details our methodology, the rationale behind our design choices, and the challenges encountered in developing a robust, real-time posture monitoring system.
We identify five crucial postures associated with neck strain: Neutral, Forward Head Posture (FHP), Slight Neck Bend, Severe Neck Bend, and Hunching as shown in the Figure \ref{fig:postures}. These postures were selected based on their clinical relevance and ability to comprehensively represent the spectrum of neck positions during device use. The neutral posture serves as a baseline, while FHP is a primary indicator of tech neck syndrome, associated with increased neck muscle strain \cite{hansraj2014assessment}. Slight and Severe Neck Bends represent progressive stages of poor posture, allowing for early intervention and tracking of posture degradation over time and hunching was included due to its impact on both neck and upper back \cite{park2015effects}. It is essential to correctly classify these postures as it requires different alerts for every posture. 

\textbf{Neck Flexion Angle using IMU}
We use IMU sensor to sense the neck angle. We explore various IMU readings like 3-axis acceleration, roll, pitch, yaw, and 3-axis displacements. We found that pitch is most correlated with the neck angle and hearable\'s IMU offer ideal position to measure neck flexion angle. 
Additionally, we use the x-axis displacement, derived from the double integration of acceleration, to track head movement as we go closer to the screen in some postures. 
As can be seen in Figure \ref{fig:key}, the pitch and displacement values are distinct for each of the 5 postures. Note, the displacement values keep going back to zero because the double integration method is unreliable, so it is programmed to go back to zero). 
This shows that we can use these IMU features to distinguish between the 5 postures. 
However, relying solely on these features presents challenges like variations in individual anatomy and movement patterns. There could be variations in individual neutral positions that would require calibration. Moreover, external factors such as device placement and sensor drift can affect the accuracy of the measurements. \\
\textbf{Distance from Screen using Microphones}
Distance from the screen is an important indicator of digital eye strain (DES). We use microphones from the hearables and speaker from the device to give distance estimate from the screen. The idea is to use the speaker from the device to emit a chirp signal, which is then received by the hearables' microphones and used to estimate the distance to the screen using Time-of-Flight (ToF) based ranging.  As shown in Figure \ref{fig:key}, when user performs these postures, they are not in the recommended distance from the screen range, which puts them at the risk of DES. 
Thus, it is essential to monitor user's distance from the screen. Additionally, this information helps in making the posture classification more robust as the distance values for the 5 postures are different as shown in Figure \ref{fig:key}. Using only audio-based distance measurements can be affected by environmental noise and reflections, leading to inaccurate readings. Additionally, the varying positions of the microphones on different headsets can introduce inconsistencies in the measurements. The subtle head movements can also affect the accuracy of distance estimation.  \\
\textbf{Why Sensor Fusion?}: NeckCare uses sensor fusion approach to that leverages the complementary strengths of IMU and audio sensing to detect posture. The IMU provides high-frequency data on head orientation and movement, while acoustic ranging offers precise distance measurements. This fusion allows for more robust posture classification and enables the detection of subtle postural changes that might be missed by a single modality. For example, hunch and severe neck bend postures can be confused if only detected by IMU sensor. Similarly, postures like FHP and slight neck bend can be confused if only sensed by acoustic ranging. Thus, the two sensor streams provide more holistic view of the posture.




\section{NeckCare System}\label{sec:system}

\subsection{NeckCare Prototype}
NeckCare prototype consists of a head tracker (HT) sensor placed on the top of a headset, and two microphones attached to it. We use an Arduino based head tracker using an ST-Micro IMU and EMW omnidirectional hardwire lavalier microphone \cite{microphones}. The microphones are connected to the Fireface UCX audio device driver \cite{website:FirefaceUCX}, which had an additional speaker connected to it. The speaker is a mini racetrack driver manufactured by Neosonica,  measuring 12mm x 40mm, housed in a custom 3D printed enclosure which is similar to loudspeakers that might be found in laptops or tablets. We use the audio device driver for prototype proposes, as we need synchronization between the speaker and microphones for ToF-based ranging. We place the speaker on the laptop to mimic that the device is emitting the chirp signal. 
\subsection{Classifying Tech Neck Postures}
\begin{figure*}[t]
    \centering
    \includegraphics[width=0.7\textwidth]{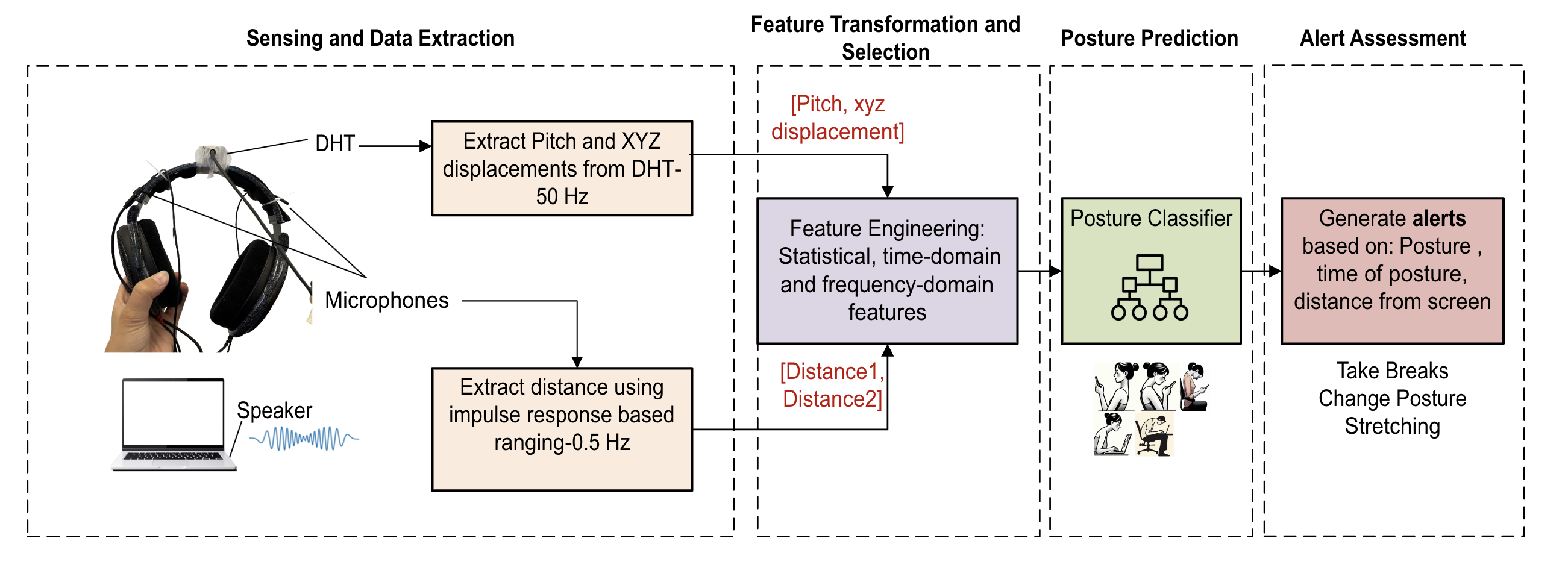}
    \caption{\textit{NeckCare} overview}
    \vspace{-0.1in}
    \label{fig:system}
\end{figure*}
NeckCare employs a novel sensor fusion approach, combining IMU data with acoustic ranging to accurately monitor neck posture. The system pipeline consists of several key components: Sensing and Data Extraction, Feature Engineering, Posture Prediction, and Alert Assessment.
The first phase involves collecting data from IMU sensors and capturing audio signals for distance estimation using microphones. The IMU sensors embedded in the hearables collect data on the pitch angle and x-axis displacement of the user's head at 50 Hz, which is crucial for determining the neck flexion angle and lateral movements. Meanwhile, the speakers on the laptop  emit chirp signals every 0.5 s and the microphones on the headset capture the audio signals to calculate the distance from the screen. 
The raw IMU data is filtered to remove noise and smooth the signals, ensuring that the pitch and displacement data are accurate and reliable for feature extraction. Similarly, the audio signals are processed to extract the ToF information, which in turn gives the distance from screen.
After the prepossessing, we get these readings: [timestamp, pitch, displacement, distance1, and distance2]. Later, we derive the statistical, time, and frequency domain features from these sensor readings, including mean, standard deviation, minimum, maximum, peak frequency, mean frequency, skewness, and kurtosis. 
We split the data into train-test split, where data from 10 participants is used for training and rest is used for testing. The training data includes features from multiple participants and corresponding posture labels. 
The classification phase involves training a Random Forest Classifier with 100 estimators using the extracted features. The trained model then classifies the user's sensor readings into one of the five identified postures. This classification helps in providing real-time feedback and alerts to the user.
Finally, the feedback component of the system continuously monitors the user's posture and provides real-time feedback to help maintain proper posture. When one of the poor posture is detected, the system generates alerts to remind the user to adjust their posture if the time in posture exceeds the threshold value, helping to prevent neck strain and promote better posture habits. The alerts can also be based if the user is very close to the screen for long time. The alerts could vary depending on individuals muscle strength, age, and working hours. We can provide alerts like take breaks, do neck stretches, or correct posture after consulting the physician.

\section{Evaluation}\label{section:eval}

\subsection{Implementation}

The head tracking system was implemented using a Python script interfacing with a custom C library. A shared C library was developed to interface directly with the head tracking hardware. The python script initializes the tracking device, then enters a continuous loop to acquire Yaw-Pitch-Roll (YPR) and XYZ positional data at approximately 50 Hz. Data is obtained by calling C functions through ctypes, stored in float arrays, and logged to a CSV file along with timestamps. 
This architecture allows for real-time data acquisition and storage, facilitating subsequent analysis of head movement patterns while providing immediate console output for monitoring. Data is logged in CSV format, with each row containing: Timestamp, YPR values (3 floats), XYZ values (3 floats). \\
For implementing acoustic distance measurement system using MATLAB, employing a 18kHz-24kHz chirp signal sampled at 48 kHz. The system utilizes a Fireface UCX \cite{website:FirefaceUCX} audio interface for precise signal transmission and dual-channel recording. For simultaneous playing and recording of sound, we use audioPlayerRecorder MATLAB library. Cross-correlation between the recorded signals and the original chirp determines the ToF, which is converted to distance using the speed of sound and corrected for system latency. We find the loopback latency by placing the microphones very close to the speaker and subtract it from the estimated distance values to get corrected distance. The script calculates distances to two microphones and log timestamps, raw and corrected distances to a CSV file. 
We merge these two sensor streams using timestamp and then use this data to train the random forest classifier using Python scikit-learn library. For real-time operation, we predict the posture using this trained model. 




\subsection{Data Collection and Evaluation metrics}
To evaluate the performance of posture classification, we collect a dataset from 15 participants (6F, 9M), each performing the five identified postures. The participants wore headset equipped with IMU sensors and microphones, and data was collected over multiple sessions to account for variations in posture and environmental conditions. We showed them multiple pictures of each posture and asked them to hold that posture for 3 mins. Each session included a calibration phase to establish the neutral position for each participant, followed by the collection of IMU and audio data for each posture. We use accuracy, precision, recall, F1, and Gini importances to understand the effectiveness of NeckCare.
Additionally, we conduct experiments to evaluate the acoustic distance estimation system using the chirp emitted at the SPL of 55dB A-weighted at a distance of 0.25 meters. We move our prototype and test at distances of 0.25m, 0.50m, and 1.00m under four conditions: silence, pink noise (75dB C-weighted), pop music (82dB C-weighted), and silence with simulated small head movements. For each scenario, we record 100 distance estimates. We use mean and standard deviation of distance measurement to evaluate the accuracy of distance estimation.

\begin{figure}[t]
    \centering
    \includegraphics[width=0.75\columnwidth]{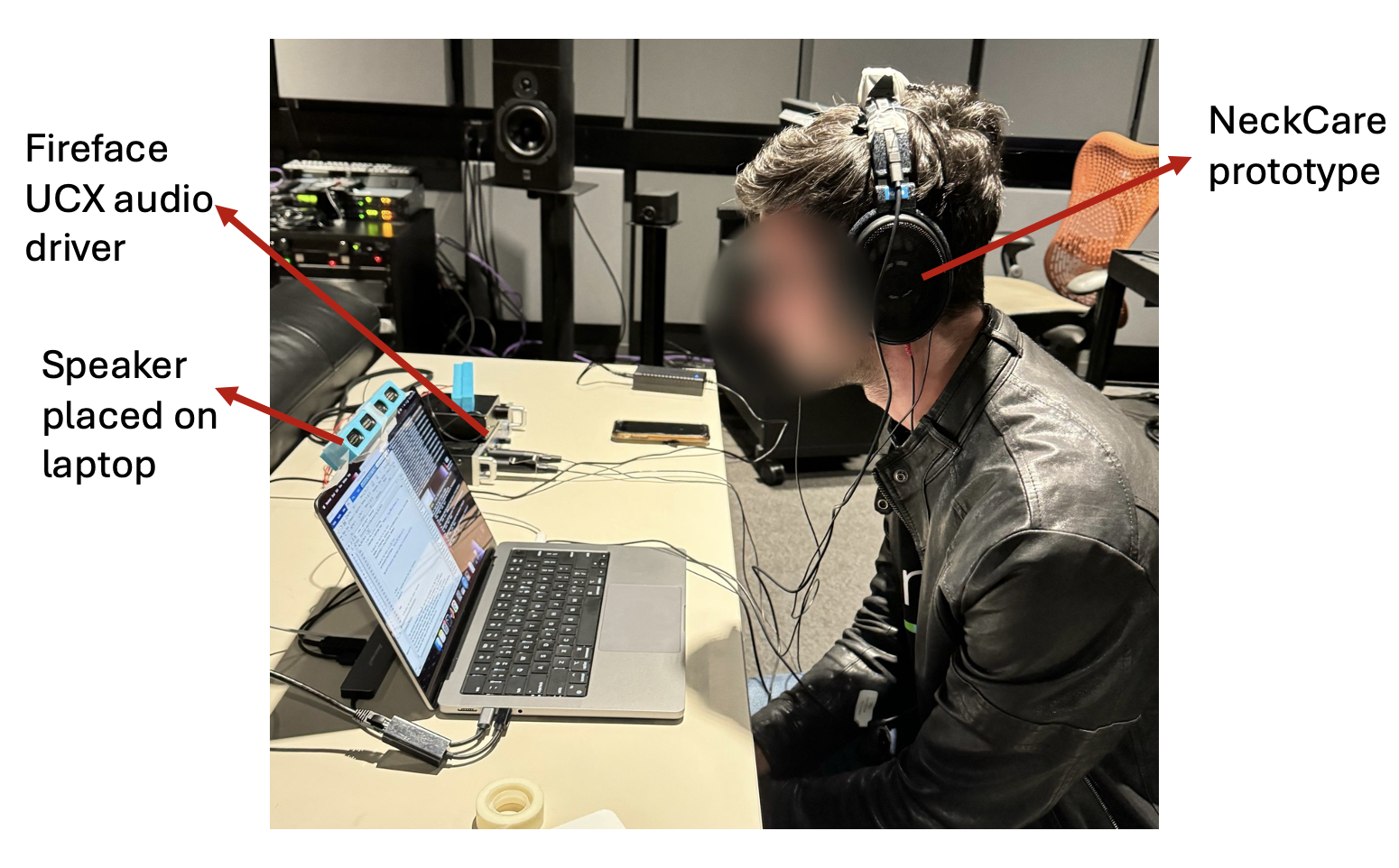}
    \caption{Data collection setup}
    \label{fig:exp}
\end{figure}

\subsection{Results}

We  demonstrate the effectiveness of NeckCare in accurately classifying neck postures and measure distance from screen in real-world conditions.

\begin{figure}[t]
    \centering
    \includegraphics[width=0.6\columnwidth]{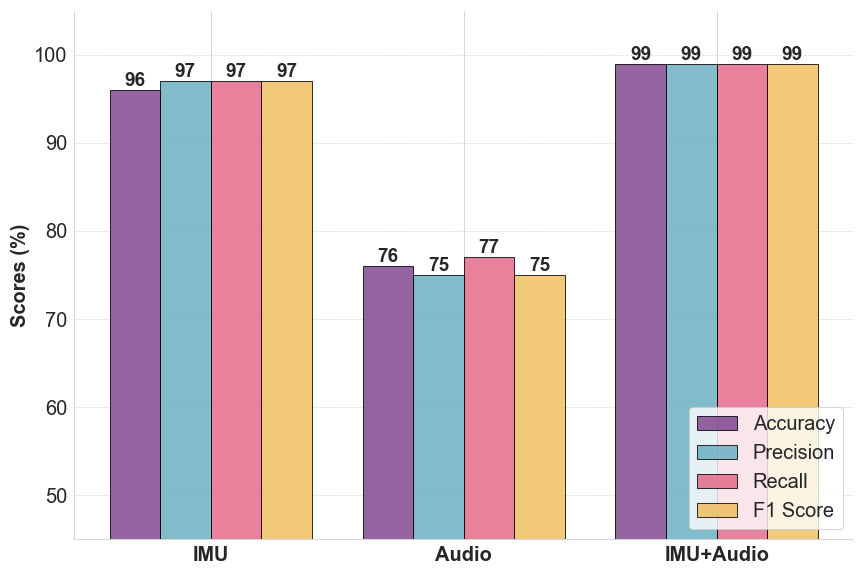}
    \caption{Posture Classification using different modalities}
    \label{fig:acc}
\end{figure}

\begin{figure}[t]
    \centering
    \includegraphics[width=0.6\columnwidth]{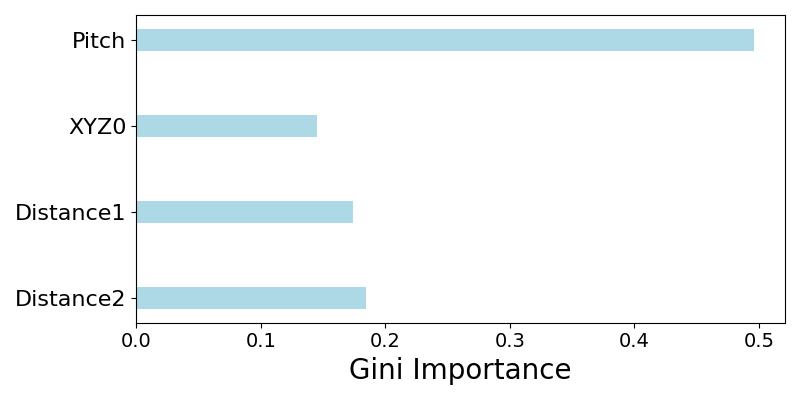}
    \caption{Feature Importances}
    \label{fig:feat}
\end{figure}

\textbf{Different Sensing Modality}: We train different Random Forest Classifiers that use only IMU, only Audio, and both features. Figure \ref{fig:acc} shows the performance of each of these three classifiers. IMU-only model achieves an high average accuracy of 96\%, where the audio-only model has low accuracy of 76\%. The combination of both sensing modalities resulted in 3\% accuracy improvement over IMU-only model. While this improvement is very less, audio only model offers additional advantages of giving distance to the screen. 
Our IMU and audio trained model is 2.9 MB and induces a prediction latency of 4 us. This means the proposed system can run in real-time on resource constrained devices.
These results indicate that NeckCare can reliably classify different neck postures with high accuracy, precision, and recall. The system’s ability to provide real-time feedback and alerts based on posture classification can help users maintain proper posture and prevent neck strain. 
Since IMU only model is very accurate by itself, it can be used as a standalone system or can be used to trigger the audio sensing based system to reduce battery consumption.

\textbf{Feature Importances}: We extract the Gini importances from the trained random forest model and found that pitch is the most important feature for posture classification, which is expected. The distance values derived from the microphones was the next important feature and both distances had similar importance. Lastly, the x-axis displacement showed slightly less importance than the distances. 

\textbf{Accuracy of Distance Estimation}: Our results reveal remarkable stability in static conditions, with consistent millimeter-level accuracy across silent and noisy environments, demonstrating the system's robustness to background noise. However, when we simulate small head movements, we observe increased variability, with estimate jumps of a few centimeters due to shifts in the impulse response's dominant peaks. Despite this, we find the system's performance remains within an acceptable range for its intended application. Our experiment highlights the system's strong noise resistance and overall accuracy, while also revealing potential areas for improvement in handling minor positional changes. By including a movement condition, we provide a more realistic assessment of real-world performance where perfect stillness is impractical. 

\textbf{Key Takeaways}: Pitch is the most important feature for detecting postures related to tech neck.
Distance from the screen is a key indicator of DES, and improves posture classification accuracy. 
NeckCare detects posture in real time with very low prediction latency and is robust to environment noise and user factors.

\begin{figure}
    \centering
    \includegraphics[width=0.9\columnwidth]{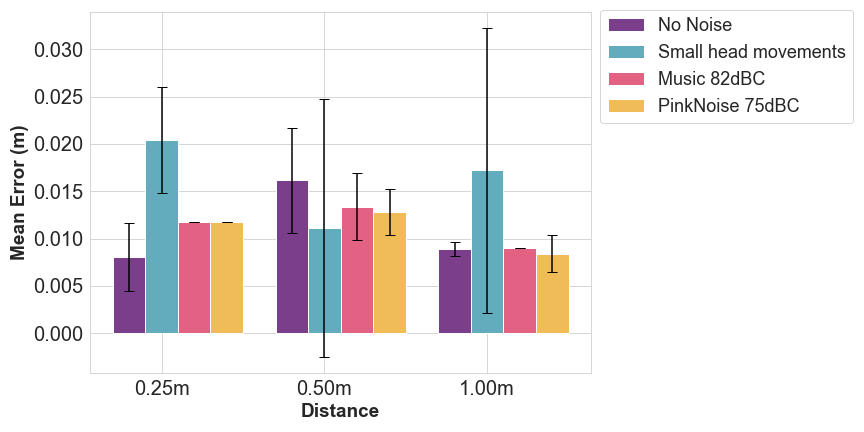}
    \caption{Distance Estimation Error for Different Conditions}
    \label{fig:enter-label}
\end{figure}













\section{Discussions}~\label{sec:discussion}
\begin{figure}
    \centering
    \includegraphics[width=0.6\columnwidth]{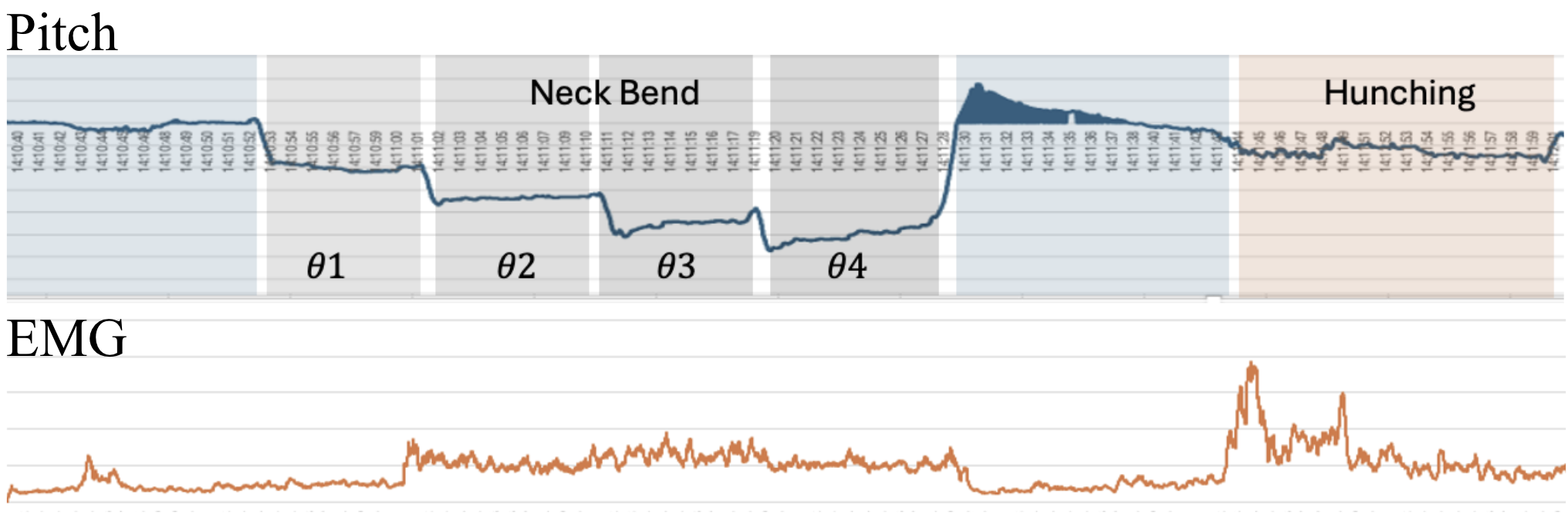}
    \caption{EMG and Pitch for different neck postures}
    \label{fig:emg}
\end{figure}
Our study on the NeckCare system revealed several challenges that could impact its accuracy and reliability. A primary concern is the variability in individual users' neutral head positions, which affects baseline measurements and necessitates frequent calibration. Additionally, differences in hearable placement and orientation can introduce inconsistencies in sensor readings. The system's requirement for constant headset wear and synchronization for screen distance monitoring also presents practical limitations. To address these issues, future work will focus on developing advanced, automated calibration techniques that can adapt to individual posture patterns, reducing the need for manual adjustments. Another limitation of this work is that it only focuses upper back posture and does not sense the entire sitting posture like lower back or shoulders. Furthermore, conducting comprehensive user studies across diverse populations will be crucial in refining the system's performance and optimizing it for real-world applications across various demographics and usage scenarios. 
Overall, we plan to extend this work in the following research directions: 
\textbf{Measuring Muscle Load}: We plan to sense the muscle load directly in addition to the posture. This will help in providing interventions when the muscle is about to get strained. For this, we use EMG sensors as the muscle load groundtruth as they can sense muscle activity and strain. We perform experiments by placing EMG electrodes at the upper trapezius muscle and found that there is strong correlation between the pitch values of IMU and EMG signal envelop for different postrues. As shown in Figure \ref{fig:emg}, the EMG readings show significant variations for different neck angles ($\theta1<\theta2<\theta3<\theta4$) and hunching.  We were able to predict EMG values using pitch with an accuracy of 93\% using Random Forest Regression model. Next steps are to predict muscle load using the NeckCare system by predicting EMG. This will help in identifying what activities are contributing to the most strain in the back, so that users can be mindful of that. \\
\textbf{Personalizing Alerts}: For the alert assessment, each individual has varying muscle strengths. We can the change alert thresholds for people with weaker back muscle. For example, for people with weaker back muscles, the alerts can be generated for lesser duration of time. This can be identified using surveys or with the help of physicians. \\
\textbf{Detecting Neck Exercises}: Certain neck exercises are good for tech neck prevention like neck stretches, chin retraction, shoulder raises, etc. In addition to providing alerts to users to perform these neck exercises, we can also detect whether they are actually performing the recommended exercises to have a closed-loop management strategy.\\
\textbf{Eye Strain Prediction}: Eye strain leads to frequent blinking and change in pupil adjustment time. We can combine the proposed system with smartglasses-based eye movement tracking. This will enhance the DES prediction and management.\\
\textbf{Long-term Progression Monitoring}: Children and young adults are facing the problem of permanent FHP or deformations. With the proposed system, we can track if there are changes in their posture, so that they can consult the doctor.

\section{Conclusion}
We presented NeckCare, a system that addresses multiple tech-related health issues using sensors embedded in hearables. By leveraging IMU sensors and microphones, our system can accurately detect and classify five key neck postures. The integration of ToF based distance estimation using audio signals further enhances the system's ability to monitor posture and detect digital eye strain. Our evaluation demonstrated that NeckCare achieves high accuracy in classifying neck postures and estimating distance from screen, making it a reliable tool for real-time posture monitoring and feedback. The system's non-invasive nature and use of ubiquitous hearable devices make it a promising solution for widespread adoption. Overall, NeckCare represents a significant step forward in enabling posture monitoring and strain prevention. Our system has the potential to promote better posture habits and reduce the risk of neck strain, contributing to improved overall well-being. With NeckCare, we can help millions of office workers prevent chronic neck pain. 

\bibliographystyle{ACM-Reference-Format}
\bibliography{output}

\end{document}